\documentclass[conference]{IEEEtran}
\IEEEoverridecommandlockouts
\usepackage{cite}
\usepackage{amsmath,amssymb,amsfonts}
\usepackage{multirow}
\usepackage{algorithmic}
\usepackage{graphicx}
\usepackage{textcomp}
\usepackage{xcolor}
\usepackage[french]{babel}
\usepackage[utf8]{inputenc}
\def\BibTeX{{\rm B\kern-.05em{\sc i\kern-.025em b}\kern-.08em
    T\kern-.1667em\lower.7ex\hbox{E}\kern-.125emX}}
\begin{document}

\title{CycleGAN Voice Conversion of Spectral Envelopes using Adversarial Weights \thanks{The research in this paper has been funded by the ANR project TheVoice: ANR-17-CE23-0025}}

\author{\IEEEauthorblockN{Rafael Ferro}
\IEEEauthorblockA{\textit{IRCAM, CNRS, Sorbonne Université} \\
\textit{STMS Lab}\\
Paris, France \\
rafael.ferro@ircam.fr}
\and
\IEEEauthorblockN{Nicolas Obin}
\IEEEauthorblockA{\textit{IRCAM, CNRS, Sorbonne Université} \\
\textit{STMS Lab}\\
Paris, France \\
nicolas.obin@ircam.fr}
\and
\IEEEauthorblockN{Axel Roebel}
\IEEEauthorblockA{\textit{IRCAM, CNRS, Sorbonne Université} \\
\textit{STMS Lab}\\
Paris, France \\
axel.roebel@ircam.fr}
}

\maketitle

\begin{abstract}
This paper tackles GAN optimization and stability issues in the context of voice conversion. First, to simplify the conversion task, we propose to use spectral envelopes as inputs. Second we propose two adversarial weight training paradigms, the generalized weighted GAN and the generator impact GAN, both aim at reducing the impact of the generator on the discriminator, so both can learn more gradually and efficiently during training. Applying an energy constraint to the cycleGAN paradigm considerably improved conversion quality. A subjective experiment conducted on a voice conversion task on the voice conversion challenge 2018 dataset shows first that despite a significantly reduced network complexity, the proposed method achieves state-of-the-art results, and second that the proposed weighted GAN methods outperform a previously proposed one.
\end{abstract}


\section{Introduction}
\label{sec:intro}

\subsection{Related works}
Voice identity conversion (VC) consists in modifying the voice of a source speaker so as to be perceived as the one of a target speaker. 
Over the past few years, VC has largely gained in popularity and in quality  \cite{Tod16, Lor18}, in particular with the development of neural voice conversion algorithms \cite{tanaka_atts2s-vc:_2018, kameoka_convs2s-vc:_2018}.
VC consists in learning a conversion function between the acoustic space of a source and a target speaker. 
This conversion function is generally learned from a pre-aligned database (parallel VC) in which the source and the target speakers pronounce the same set of sentences, so that a direct correspondence between the frames of the source and target speakers can be established. Unfortunately, this constraint reduces the amount of available recordings of source and target speakers. \\
\indent Modern VC is mainly based on neural architectures with the particular objective to extend the VC from parallel to non-parallel speech databases. The main advantage of non-parallel VC is that it provides the flexibility to learn the conversion from "on-the-fly" speech databases, which can more easily handle large amount of data and accommodate multiple speakers. These architectures includes Variational Autoencoders (VAEs) \cite{kameoka_acvae-vc:_2018, hsu_voice_2017, huang_refined_2018}, Generative Adversarial Networks (GANs) \cite{kameoka_stargan-vc:_2018, kaneko_stargan-vc2:_2019, kaneko_parallel-data-free_2017, kaneko_cyclegan-vc2:_2019, fang_high-quality_2018}, Phonetic PosteriorGrams (PPGs) \cite{sun_phonetic_2016} and sampleRNNs \cite{Zhou2018} among others. 
%
%
The use of GAN architectures \cite{goodfellow_generative_2014} for VC is inspired by advances conducted in the fields of image generation and manipulation. The cycleGAN is a particular configuration of GANs which has been specifically formulated to learn transformations between two different domains or between unaligned or unpaired datasets with application to image-to-image translation \cite{zhu_unpaired_2017}.  \\
\indent The cycleGAN-VC is the extension of the cycleGAN to the VC task, which has become a standard in non-parallel VC \cite{kaneko_parallel-data-free_2017, kameoka_stargan-vc:_2018}. The main idea behind the application of the cycleGAN to VC is that the cycle-consistency encourages the preservation of the phonetic content through the cycle while learning to modify the speaker identity. 
Despite the advances accomplished in non-parallel VC, cycleGAN-VC still suffers from important limitations which conduct to conversion of mitigated quality. One main limitation is due to the well-known stability issues of the GAN \cite{salimans_improved_2016, radford_unsupervised_2015}. This issue, combined to the limited amount of data available in the VC task, can lead to severe degradation of the voice conversion quality and the naturalness of the converted speech. 

Tackling GAN stability issues has motivated the development of multiple ideas and heuristics \cite{salimans_improved_2016}, such as the popular Wasserstein GAN \cite{arjovsky_wasserstein_2017} or the LSGAN \cite{mao_least_2016}. More recently, so as to tackle stability issues, the weighted GAN has been introduced \cite{pantazis_training_2018}. This novel approach, based on game theory, instead of equally weighted “fake” samples, more attention is given to samples that fool the discriminator. A particular form of weighted GAN has been recently applied to VC. \cite{paul_non-parallel_2019}. 

\subsection{Contribution of the paper}

This  paper  proposes two contributions. First, we propose to use spectral envelopes as inputs instead of using cepstral coefficients as in \cite{kaneko_parallel-data-free_2017}. The use of the spectral envelope is assumed to simplify the task for the convolutional networks that will be required to only slightly move the spectral formants. Thus we assume to not need an extremely deep network and we will show that we can achieve similar performance with a significantly smaller network. Second, we propose to introduce a constraint as an additional training loss that enforces to preserve energy contour of the converted speech signal. This allows us to add a generator loss not depending on adversarial training, which is expected to contribute to the stability of the training. Third, we present a novel method to tackle the stability issue of GAN training, exploring a novel weighted GAN approach. We achieve this by adding a weight to the loss of the discriminator, giving more weight to “true” samples rather than to “fake” ones.

%

\section{Preliminary Works on CycleGAN VC}
\label{sec:GAN}

\subsection{Generative Adversarial Networks}
A Generative Adversarial Network (GAN) \cite{goodfellow_generative_2014} is a neural network system composed by a generator G and a discriminator D, in which the discriminator is trained to discriminate real samples from generated samples, while the generator is trained to generate real-like samples, using the discriminator as the decision rule. The objective can be written as:
\begin{equation}\label{eq1}
\begin{split}
\min_{G} \max_{D} \mathcal{L}(D,G) = & \mathbb{E}_{x \sim p(X)}[\log D(x)]  \\
+ & \mathbb{E}_{z \sim p(Z)}[\log(1-D(G(z)))]
\end{split}
\end{equation}
where: $x$ is a sample from a distribution $p(X)$ to be modeled, $z$ is a sample generated from a random distribution $p(Z)$, and $\mathbb{E}_{x \sim p(X)} $ represents the expected value of $x$ given the distribution $p(X)$.

\subsection{Cycle Generative Adversarial Networks}
In the cycleGAN architecture \cite{zhu_unpaired_2017, yi_dualgan:_2017}, a generator $G_{X\rightarrow Y}$  reads data from a dataset X and learns to map it into its respective position in a dataset Y, and vice versa for a generator $G_{Y\rightarrow X}$. If X and Y represent languages, this system should be analogous to two translators. To train these generators, the cycleGAN framework uses two adversarially trained discriminators to discriminate respectively any $x \in X$ in relation to $G_{Y\rightarrow X}(y)$ for any $y \in Y$  and any $y \in Y$ in relation to $G_{X\rightarrow Y}(x)$ for any $x \in X$. Since $G_{Y\rightarrow X}(G_{X\rightarrow Y}(x))$ should be equal to x, and $G_{X\rightarrow Y}(G_{Y\rightarrow X}(y))$ should be equal to y, a loss named cycle-consistent loss is added to enforce this constraint. In the following equations we state the total objective of the cycleGAN, where $\mathbb{E}_{y \sim P_{Data}(y)} $ represents the expected value for the distribution Y and $\mathbb{E}_{Y \sim P_{Data}(X)} $ represents the expected value for the distribution X.

The following equation describes the adversarial loss for the discriminator $D_{Y}$ (the equation for $D_{X}$ is analogous):
\begin{equation} \label{eqE2}
\small
\begin{split}
\mathcal{L}_{adv}(G_{X\rightarrow Y},& D_{Y}) = \mathbb{E}_{y\sim P_{Data}(y)}[\log(D_{Y} (y))]  \\
 +&  \mathbb{E}_{x\sim P_{Data}(x)}[\log(1-\phantom{} D_{Y}(G_{X\rightarrow Y}(x)))].
\end{split}
\end{equation}

The following equation describes the cycle-consistency loss, using L1 norm:
\begin{equation} \label{eqE3}
\small
\begin{split}
\mathcal{L}_{cyc}(G_{X\rightarrow Y}, & G_{Y\rightarrow X}) = \\
& \mathbb{E}_{x\sim P_{Data}(x)}[||G_{Y\rightarrow X}(G_{X\rightarrow Y} (x)) - x||_{1}]  \\
+ &  \mathbb{E}_{y\sim P_{Data}(y)}[||G_{X\rightarrow Y} (G_{Y\rightarrow X} (y)) - y||_{1}].
\end{split}
\end{equation}

The following equation describes the total objective of the cycleGAN, where \(\lambda_{c}\) represents the weight for the cycle-consistency loss:
\begin{equation} \label{eqE4}
\small
\begin{split}
\mathcal{L}_{full} =& \mathcal{L}_{adv}(G_{X\rightarrow Y}, D_{Y}) + L_{adv}(G_{Y\rightarrow X}, D_{X}) \\
+ &  \lambda_{c}L_{cyc}(G_{X\rightarrow Y}, G_{Y\rightarrow X})
\end{split}
\end{equation}

The cycleGAN-VC, introduced by \cite{kaneko_parallel-data-free_2017}, is trained to convert a source speaker Mel Frequency Cepstral Coefficients (MFCCs) into a target speaker MFCCs, so as to perform VC. Their discriminators task is therefore to discriminate whether the conversions belong to their respective target speaker identity or not. In particular, so as to adapt the original cycleGAN framework, they used Gated CNNs as well as an identity-mapping loss, which is reported to encourage phonetic invariance. However, we did not find these two additional ideas beneficial. 

\section{GAN with Adversarial Weights}
\label{sec:optimization}

Generative Adversarial Networks are reported to be difficult to train. One problem are vanishing gradients when the discriminator achieves perfect discrimination, or when the generator is able to perfectly fool the discriminator, though it is producing nonsense. Another problem is the instability that is due to the fact that the discriminator is trained to systematically reject generated examples independent of their quality. In the case when the generator generates target samples covering only a small part of the target space the discriminator will improve its objective by means of pushing the generator out of the target space even if it has to wrongly classify some of the real samples as well. As a result the discriminator will push the generator away from the target space hindering the generator to converge.

To solve these issues, many ideas have been proposed, such as the DCGAN architecture \cite{radford_unsupervised_2015}. \cite{gulrajani_improved_2017} discussed mini-batch discrimination, historical averaging, one-sided label smoothing and virtual batch normalization. Also, new losses have been proposed, such as the Wasserstein GAN \cite{arjovsky_wasserstein_2017} and the LSGAN \cite{mao_least_2016}.

\subsection{Weighted GAN System}

\begin{figure*}[ht!]
\centering
\includegraphics[width=0.7\columnwidth]{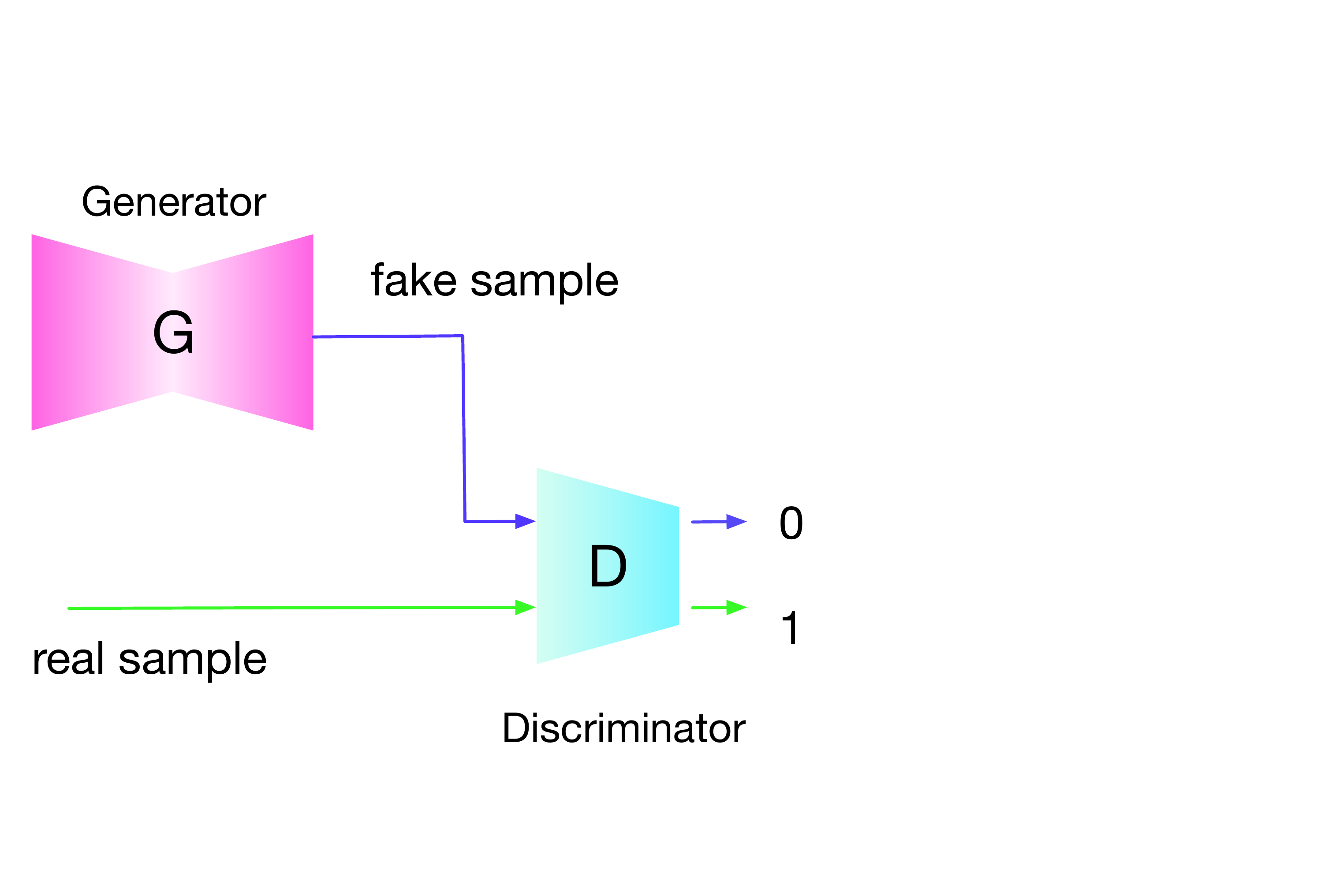}
\hspace{1,5cm}
\includegraphics[width=0.7\columnwidth]{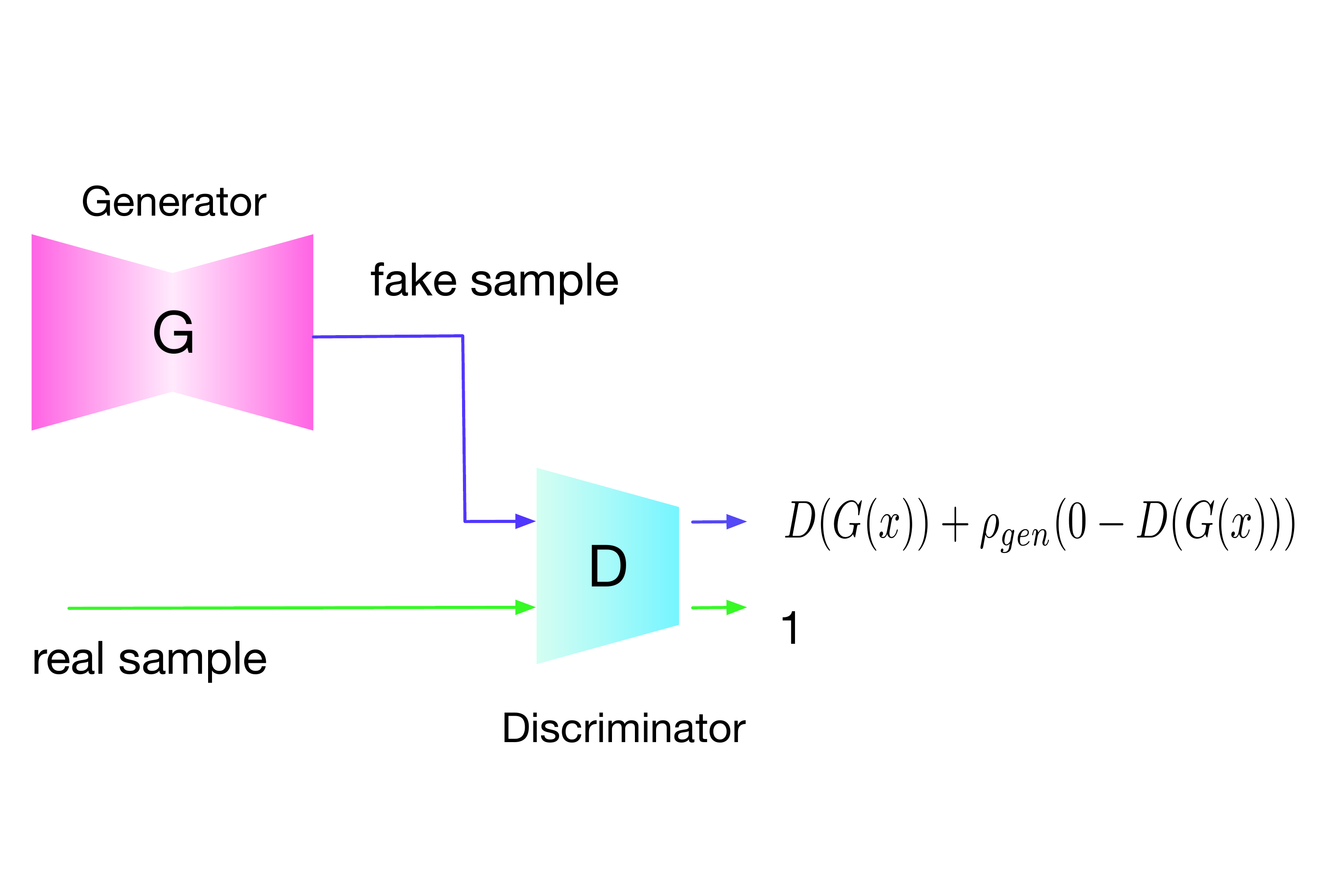}
\vspace{-0.5cm}
\caption{Comparison of GAN with and without generator impact. On left: vanilla GAN, on right: proposed GAN with weighted labels.}
\label{fig:sotfGAN}
\end{figure*}

Recently, so as to tackle CycleGAN optimization, Paul et al. developed the weStarGAN \cite{paul_non-parallel_2019}, implementing the weighted GAN idea \cite{pantazis_training_2018} to the starGAN-VC architecture \cite{kameoka_stargan-vc:_2018}. They do so by multiplying sample-wise a coefficient to the generator loss. Their idea is to give less weight to samples poorly produced by the generator, so that the generator has a stronger motivation to produce samples similar to the real data.  For each sample $j$ they compute its respective normalised weight $w_{j,g}$, while introducing a hyper-parameter $\eta_{gen}$:
\begin{equation} \label{eqE5}
\small
\begin{split}
w_{j,g} = {e^{\eta min(0, D_{j})}} \\
w_{j,g} = \dfrac{w_{j,g}}{\sum_{j=1}^{m} w_{j,g}}, j = 1, ..., m.
\end{split}
\end{equation}

This coefficient is then multiplied sample-wise by the generator loss. Note that samples that the discriminator does not correctly discriminate have a higher weight in the generator loss. A hypothesis assumed in \cite{paul_non-parallel_2019} is that the discriminator is “faithful”: that quantitatively it returns on average values above 0.5 when the samples come from the real distribution and below 0.5 when fake samples are fed to the Discriminator. This structure is our starting point in our research on the stability and optimization of the GAN.

\subsection{Generalized Weighted GAN}

Since in practice the discriminator might not be “faithful”, in order to encourage its “faithfulness”, we propose to train the discriminator rather by fairly produced samples than by poorly produced ones. This helps to reduces the above mentioned instability caused by training the discriminator to judge generated data, no matter how good it is, as {\it equally} wrong. To do so, we further develop Paul et al. idea, by also applying weights to the discriminator loss. We followed the exact same procedure: we introduce a hyper-parameter $\eta_{dis}$ and for each sample we compute its respective normalised weight $w_{j,d}$, following the same procedure as in Eq. (\ref{eqE5}). This coefficient is then multiplied sample-wise by the discriminator loss. Thus samples perceived as poor by the discriminator have less weighting in the discriminator loss. We name this system as generalized weighted GAN (geweGAN).

\subsection{Generator Impact GAN}

Finally, we introduce a novel weighted GAN technique, also encouraging the discriminator “faithfulness”. In the proposed approach, the idea is to encourage the training of the discriminator by weighting samples produced by the generator by a constant generator impact hyperparameter $\rho_{gen}$, so that the discriminator is rather encouraged by the ‘‘real" distribution than by the generated one. We name this system the generator impact GAN (gimGAN), since we can control the impact of the generator on the discriminator loss. 
The idea is equivalent to replacing the ‘‘hard" label 0 scored by the discriminator on the samples produced by the generator $D(G(x))$, by a ‘‘soft'' label calculated as:
\begin{equation}
\small
D(G(x)) + \rho_{gen}(0 - D(G(x)))   \in \left[0, 1\right]
\end{equation}
where: $\rho_{gen}(0 - D(G(x)))$ represents the current distance of the discriminator relatively to the target ‘‘hard'' label 0, weighted by generator impact hyperparameter $\rho_{gen}$.
Once again, the motivation behind is to tackle the fact that in the vanilla GAN, the discriminator is trained to consider samples produced by the generator as ‘‘fake'' ones, whatever its quality is, which might have the effect to discourage the generator to converge to the correct distribution. The principle is illustrated in Figure \ref{fig:sotfGAN}.


\subsection{Energy constraint}

Additionally, we further apply an energy constraint to the cycleGAN architecture. This constraint enforces preservation of the energy contour of the original source speech signal during conversion, and avoids incoherence between source and converted envelope in the converted speech signal. Furthermore, by means of providing stable feedback to the generator, this constraint is expected to reduce the instability of the GAN training. To achieve this, we impose a reconstruction loss on the amplitude mean for each frame, on both generators. Since we work with spectral envelopes as inputs, we just add the following term to the total loss, where \(\lambda_{c}\) represents the weight for the cycle-consistency loss:
\vspace{-0.2cm}
\begin{equation} \label{eqE6}
\small
\begin{split}
\mathcal{L}_{e} =& \lambda_{e}\mathbb{E}_{x\sim P_{Data}(x)}[||\sum_{t=0}^{T}G_{X\rightarrow Y}(x)-x||_{1}] \\
&+ \lambda_{e}\mathbb{E}_{x\sim P_{Data}(x)}[||\sum_{t=0}^{T}G_{Y\rightarrow X}(y)-y||_{1}]
\end{split}
\end{equation}

This constraint enforces preservation of the energy contour of the original source speech signal during conversion, and avoids incoherence between source and converted envelope in the converted speech signal. Further, by means of providing stable feedback to the generator, this constraint is expected to reduce the instability of the GAN training.

\begin{table*}[!]
\caption {\label{tab:results_all}MOS and 95\% confidence interval obtained for the different VC systems.}
\centering
\begin{tabular}{|l|l|l|l|l|l|l|}
\hline
                                     & \multicolumn{2}{c}{Male-to-Male (MTM)} & \multicolumn{2}{c}{Female-to-Female (FTF)} & \multicolumn{2}{c|}{TOTAL} \\
Speech Signal Class                  & Similarity   & Naturalness & Similarity   & Naturalness & Similarity   & Naturalness\\ \hline\hline
orig: target                         & 4.96 $\pm$ 0.09 & 4.91  $\pm$0.12 & 5.00 $\pm$0.00 & 5.00  $\pm$0.00 & 4.98 $\pm$0.04 & 4.96  $\pm$0.05 \\ \hline
conv: weGAN                          & 2.52 $\pm$ 0.42 & 2.15  $\pm$0.23  & 2.29 $\pm$0.40 & 2.13  $\pm$0.34 & 2.42 $\pm$0.30 & 2.16  $\pm$0.20  \\ \hline
conv: geweGAN                        & 2.80 $\pm$ 0.76 & 2.20  $\pm$0.30 & \textbf{3.32 $\pm$0.35} & 3.00  $\pm$0.40  & 3.10 $\pm$0.47 & 2.56  $\pm$0.28\\ \hline
conv: gimGAN                         & 3.21 $\pm$ 0.37 & 2.19  $\pm$0.42 & 3.09 $\pm$0.46 & \textbf{3.41  $\pm$0.36} & \textbf{3.15 $\pm$0.30} & \textbf{2.81  $\pm$0.33} \\ \hline
conv: cycleGAN-VC (baseline)         & \textbf{3.33 $\pm$ 0.46} & \textbf{2.43  $\pm$0.42} & 2.75 $\pm$0.33 & 2.06  $\pm$0.28  & 3.05 $\pm$0.31 & 2.27  $\pm$0.27  \\ \hline
\end{tabular}
\end{table*}

\section{Experiment}
\label{sec:experiment}

The proposed CycleGAN architectures have been trained and evaluated, using the VCC2018 database \cite{Lor18}. The VCC2018 training corpus contains 80 short sentences per speaker, sampled at 16 kHz and quantified on 16 bits. For the evaluation set, we used the first 5 sentences, whose length was superior to 2s.

Our architecture was inspired by the DCGAN and by the cycleGAN-VC \cite{kaneko_parallel-data-free_2017, kaneko_cyclegan-vc2:_2019}. For both generators, we used a two-layer encoder with convolutions, followed by a two-layer bottleneck with convolutions and then a  two-layer decoder with transposed convolutions. For both the encoder and the decoder, we applied a kernel size 2 and a stride size 2, so as to avoid the checkerboard effect \cite{odena2016deconvolution}, noticing that consistently better results were obtained when this undesirable effect was avoided. For the bottleneck, we applied kernel size 3 and and stride size 1. The overall generator has respectfully 256, 512, 512 , 512, 256 and 1 filters. We applied instance normalization, followed by a ReLU at the end of each layer. Since spectral envelopes explicitly contain formant information, the task becomes much easier allowing us to implant a rather small generator, compared to traditional cycleGAN implementations.  \cite{kaneko_parallel-data-free_2017, kaneko_cyclegan-vc2:_2019} For both discriminators, we used 4 convolutional layers with a filter size 2, a kernel size 2, with respectfully 64, 128, 256 and 512 filters. These four layers were followed by two fully connected layers, with 512 and 1 neurons respectively. We applied instance normalization, followed by a LeakyReLU at the end of each layer, except for the last one. Our inputs were of size 32 and 128, for frequency bins and time frames respectively. We chose \(\lambda_{c}\) and \(\lambda_{e}\) to respectfully be 0.3 and 1. By lowering the traditional values for the cycle-consistency loss, we rather force GAN learning than cycle-consistency learning. This means that more weight is given to the identity learning, the task of the GAN, than to the phoneme reconstruction learning, the cycle-consistency task. In fact, otherwise results were found to be closer to a reconstruction, rather than a conversion.  We applied a batch size 1 and we used least squares error for the discriminator loss, introduced in the LSGAN \cite{mao_least_2016} and optimized it with the Adam algorithm, as in \cite{kaneko_parallel-data-free_2017, kaneko_cyclegan-vc2:_2019} for a total of 800k iterations, with a generator learning rate of 0.0002 and a discriminator learning rate of 0.0001.
We implemented three  GAN variations that all share the same network structure and input representation. The first variation, uses the Weighted GAN paradigm (weGAN) from \cite{paul_non-parallel_2019} with $\eta_{gen} = 0.1$. For the second variation, we implemented the generalized weighted GAN paradigm (geweGAN) for both the discriminator and the generator losses, as discussed above, with $\eta_{gen} = 0.9 $and $\eta_{dis} = 0.9$. For the third variation, we implemented the generator impact weighted GAN paradigm (gimGAN), as discussed above, with $\rho_{gen} = 0.9$. Finally, we also implemented cycleGAN-VC \cite{kaneko_parallel-data-free_2017} as a baseline.

\vspace{-0.1cm}

Similarly to previous research on VC, the proposed VC is focused on spectral voice conversion only. The VC is based on a source/filter decomposition of the speech signal, in which the excitation of the source speaker is preserved during conversion and only the spectral envelope conversion is learned and modified. The analysis/synthesis engine relies on superVP, an extended phase vocoder developed by IRCAM \footnote{\url{www.forumnet.ircam.fr/product/supervp-max-en/}}. The spectral envelope is estimated from the short-term Fourier transform (STFT) by using  the True Envelope algorithm \cite{Roe07}. The Mel spectral envelope is then computed by integrating the estimated spectral envelope over 32 Mel filters in which the energy of each Mel filter is normalized to unity.

\subsection{Experimental setups}

The experiment consisted into the judgment by listeners of singing voice samples, based on the similarity to the target singer and the naturalness of the singer, as used for the voice conversion 2018 challenge \cite{Lor18}. Conversion were processed for all sentences contained in the test set. For the perceptual experiment, short excerpts were used and presented to the participants (around 5s.). We chose SF4 and TM4 as the source and TF3, TF4, TM3 and TM4 as the target. We evaluated the naturalness and speaker similarity of the converted samples, with a mean opinion score (MOS) test. 

During the experiment, 15 short speech samples, original source and target speakers, and converted source-to-target speaker (each having duration of about 5s) were randomly selected from the test set, and presented to the participant in a random order. For each speech sample, the participant has the possibility to listen to an excerpt of the original target speaker. Then the participant is asked to rate the naturalness of the converted speech sample and its similarity to the target speaker. The experiment was conducted on-line, encouraging the use of headphones and quiet environment. 15 individuals participated in the experiment.

\section{Results and Discussion}
\label{sec:discusson}



The results of the perceptual evaluation are presented in table \ref{tab:results_all}. An overall result is that the original target speaker is consistently qualified to have high similarity and quality. In looking into the average results  for all speaker pairs it is easy to see that the gimGAN and geweGAN variants perform significantly better than weGAN variant. This shows that the proposed GAN modifications used for training geweGAN and  gimGAN have a positive impact on similarity and naturalness. Note that the energy constraint had an important contribution to these results. Without this constraint the similarity and naturalness rating are about 0.5 points lower (results not shown).  
Finally, the cycleGAN-VC baseline achieves approximately the same performance in similarity, while gimGAN and wegeGAN are perceived as more natural with gimGAN achieving overall the best results where differences in naturalness are more pronounced and differences in similarity remain marginal.
Note however, that the geweGAN and gimGAN networks achieve this performance with about 3 times less parameters, which results in considerably shorter training and inference times, which is expected to be a result of directly working on the spectral envelope. 
Looking into the sub groups for MTM and FTF conversions,  one can notice that the best network changes with gender. While cycleGAN-VC baseline is best for MTM conversion, wegeGAN has been evaluated best in the FTF case. These differences in the two gender groups indicate a relatively strong dependency of the conversion performance on the speaker.  

\section{Conclusion}

The present paper investigates Voice Conversion with Generative Adversarial Networks (GAN), particularly with the cycleGAN paradigm, addressing optimization and stability issues. First, we use spectral envelopes as inputs. Second, so as to optimize and to address the stability issues of the GAN training we propose a generalization of the weighted GAN, the geweGAN, and a similar approach, the gimGAN. Our conducted experiment shows first that both proposed methods are able to score a better performance than the previously proposed weighted GAN. Second, it shows that the proposed method performs similarly to the cycleGAN-VC baseline on similarity and considerably outperforms it on quality. Furthermore, these results were achieved using a significantly smaller network, which significantly reduces training time. Finally, an additional energy constraint to the loss was found to be essential for similarity and naturalness learning. For future work, we plan to combine the proposed generalized weighted GAN with the generator impact GAN, so as to further improve the stability of the GAN training procedure.

\label{sec:conclusion}

\bibliographystyle{IEEEbib}
\bibliography{strings,refs}

\end{document}